# Glial activation in white matter following ischemia in the neonatal P7 rat brain

Valérie Biran[1,2], Luc-Marie Joly[1], Anne Héron[3], Agnès Vernet[1], Céline Véga[1], Jean Mariani[1], Sylvain Renolleau[1,4] and C. Charriaut-Marlangue[1,#]

[1]UMR-CNRS 7102, Université Pierre et Marie Curie, 9 quai St-Bernard, 75005 Paris
[2]Service de Néonatologie, Hôpital Armand Trousseau, 75012 Paris
[3]Université René Descartes, 75006 Paris, France
[4]Service de Réanimation, Hôpital Armand Trousseau, 75012 Paris.

Text : 5960 words

Tables : 2

Figures : 7

[#]Correspondence :

C. Charriaut-Marlangue
UMR-CNRS 7102, HI CD, case 14
9 quai St-Bernard
75005 Paris, France.
Tel : 33 1 44 27 32 32
Fax : 33 1 44 27 22 80
Email : Christiane.Marlangue@snv.jussieu.fr

Support: The Foundation Cino del Duca supported this work.

ABSTRACT

This study examines cell death and proliferation in the white matter after neonatal stroke. In post-natal day 7 injured rat, there was a marked reduction in myelin basic protein (MBP) immunostaining mainly corresponding to numerous pyknotic immature oligodendrocytes and TUNEL-positive astrocytes in the ipsilateral external capsule. In contrast, a substantial restoration of MBP, as indicated by the MBP ratio of left-to-right, occurred in the cingulum at 48 (1.27 ± 0.12) and 72 (1.30 ± 0.18, p<0.05) hours of recovery as compared to age-matched controls (1.03 ± 0.14). Ki-67 immunostaining revealed a first peak of newly-generated cells in the dorsolateral hippocampal subventricular zone and cingulum at 72 hours after reperfusion. Double immunofluorescence revealed that most of the Ki-67-positive cells were astrocytes at 48 hours and NG2 pre-oligodendrocytes at 72 hours of recovery. Microglia infiltration occurs over several days in the cingulum and a huge quantity of macrophages reached the subcortical white matter where they engulfed immature oligodendrocytes. The overall results suggest that the persistent activation of microglia involves a chronic component of immunoinflammation, which overwhelms repair processes and contributes to cystic growth in the developing brain.

INTRODUCTION

Neonatal hypoxic-ischemic (HI) injury is a leading cause of periventricular leukomalacia (PVL), a prominent lesion of the periventricular white matter (PWM) in premature newborns. The heightened susceptibility of PWM damage stems from increased vulnerability of immature oligodendrocytes (OLs) to free radicals, cytokines and glutamate (Back *et al.*, 1998; Fern & Moller, 2000; Follett *et al.*, 2000; Volpe, 2001; Back *et al.*, 2002) and their propensity for induction of apoptosis (Han *et al.*, 2000; Puka-Sundvall *et al.*, 2000). Presumed loss of oligodendrocytes is a hallmark of PVL that results in hypomyelination and neurological deficits (Rorke, 1992; Volpe, 1997).

White matter cell loss and pyknotic immature oligodendrocytes have recently been reported in P7 (post-natal day 7) rats (Jelinski *et al.*, 1999; Follett *et al.*, 2000; Ness *et al.*, 2001) and in P9-10 mice (Skoff *et al.*, 2001) following hypoxia-ischemia. Loss of myelin basic protein (MBP) was also demonstrated 5 days after injury with substantial restoration 2 weeks later (Liu *et al.*, 2002), suggesting recovery of injured and/or formation of new OLs. Although the neonatal brain undergoes massive cell death and atrophy the first week after injury, it retains the potential to generate new oligodendrocytes up to 4 weeks within and surrounding the infarct (Zaidi *et al.*, 2004). All these data have been demonstrated in the "Rice-Vannucci model" of transient focal hypoxia-ischemia in the neonatal P7 rat, which produces infarct in the cerebral cortex, striatum, thalamus, and hippocampus and also in the white matter (Towfighi *et al.*, 1991; Towfighi *et al.*, 1994; Towfighi *et al.*, 1995). We recently developed a model of neonatal stroke, elicited by middle cerebral artery (MCA) electrocoagulation and transient homolateral carotid occlusion in 7-day-old rats (Renolleau *et al.*, 1998). This model induces ipsilateral cortical injury with apoptotic features (Renolleau *et al.*, 1998; Aggoun-Zouaoui *et al.*, 2000; Benjelloun *et al.*, 2003a). Additional neuropathological features of the ischemic lesion include acute and sustained inflammatory responses (Benjelloun *et al.*, 1999) and nitric oxide production (Coeroli *et al.*, 1998) that leads to extensive neuronal loss and the evolution of a cortical cavitary infarct (Joly *et al.*, 2003). These two models in P7 rats represent two different brain insults (hypoxia-ischemia and stroke, respectively) and can be considered complementary.

Based on our previous findings, we first investigated the process of myelination after ischemia with reperfusion to study white matter damage. At P7, oligodendrocytes are generated in large numbers in cerebrum (Skoff *et al.*, 1994). We then focused on oligodendroglial progenitors and on OLs outcome between 1 day and 2 weeks of recovery, in an attempt to study the relationship between repair processes and the presence of sustained activated microglia.

MATERIAL AND METHODS

Perinatal ischemia. All animal experimentation was conducted in accordance with the French and European Community guidelines for the care and use of experimental animals. Ischemia was performed in 7 day-old rats (17-21 g) of both sexes, as previously described (Renolleau *et al.*, 1998), minimizing the number of animals used and their suffering. Rat pups were anesthetized with an intraperitoneal injection of chloral hydrate (350 mg/kg). The anaesthetized rat was positioned on its back and a median incision was made in the neck to expose the left common carotid artery (CCA). The rat was then placed on the right side and an oblique skin incision was made between the ear and the eye. After excision of the temporal muscle, the cranial bone was removed from the frontal suture to a level below the zygomatic arch. Then the left middle cerebral artery, exposed just after its appearance over the rhinal fissure, was coagulated at the inferior level of the cerebral vein. After this procedure, a clip was placed to occlude the left common carotid artery. Rats were then placed in an incubator to avoid hypothermia. After 50 min, the clip was removed. Carotid blood flow restoration was verified with the aid of a microscope. Both neck and cranial skin incisions were then closed. During the surgical procedure, body temperature was maintained at 37-38°C. After recovery, pups were transferred to their mothers. The same surgery was performed in sham-operated rats but the left MCA and the CCA were not occluded.

Tissue preparation. Rats were anaesthetized with chloral hydrate (400 mg/kg, i.p.) and were either perfused through the left ventricle with heparinized saline followed by phosphate-buffer (PB 0.12 M, pH 7.4) containing 4 % paraformaldehyde (PFA), or killed and brains rapidly removed on an ice-cold plate. Removed brains were kept for 2 to 24 hours in the same fixative solution and placed in 0.12 M PB containing 20 % sucrose for 2-3 days. The cryoprotected brains were frozen in isopentane (-40°C) and stored at – 70°C until used. Serial coronal cryostat sections (20 $\mu$m thick) were collected on gelatin-coated slides. Some sections were stained with cresyl-violet. The cingulum and external capsule at the level of the anterior commissure (Bregma 0.2 mm) and dorsal hippocampus (Bregma – 2.5 mm) were examined in a minimum of 4-5 animals at each time point (from 12 hours to 15 days after recovery).

Measurement of infarct volume. Rats were killed at 48 hours of recovery and brains were fixed 2 days in 4 % buffered formaldehyde followed by 3 days in 20 % sucrose. Fifty-micrometer coronal brain sections were cut on a cryostat and collected on gelatin-coated slides. Sixteen sections from anterior striatum to posterior hippocampus were selected and taken at equally spaced 0.5-mm intervals. Infarct size was determined on cresyl violet-stained sections, using an image analyzer (ImagePro) as previously described (Ducrocq $et$ $al.$, 2000). The data are expressed as mean volume in mm$^3$ ± SEM.

Fluorescence $in$ $situ$ labeling of fragmented DNA. Sections were processed for DNA strand breaks (TUNEL assay) using the $in$ $situ$ Cell Death Detection Kit, Fluorescein (Roche, Meylan, France) according to the manufacturer's instructions.

<u>Immunocytochemistry</u>. The primary antibodies were directed against various antigens specific of cell types : mouse monoclonal MBP (Chemicon, MAB382, dilution 1:1000), O4 (Chemicon, clone-81, dilution 1:200) and NG2 (Chemicon, AB5320, dilution 1:200) to label mature, immature and pre-oligodendrocyte, respectively; Cy3-conjugated GFAP (Sigma, clone G-A-5, dilution 1:500) to stain astrocytes, or proliferative marker (Ki-67 rabbit polyclonal, Novocastra, NCL-Ki67p, dilution 1:200) and specific lectin (FITC-conjugated Isolectin B4, Sigma, dilution 1:500) used to label microglia. The sections, at bregma – 2.5 mm, were first incubated for 30 minutes with 5 % normal horse serum in PBS with 0.5 % triton X-100 (PBS-TX-NHS) and overnight at room temperature with appropriately diluted primary antibody in (PBS-TX-NHS), except for Ki-67 (48 hours incubation at 4°C). The primary antibodies were visualized after incubation with the appropriate species-specific biotinylated secondary antibody (Vector laboratories, AbCys, Paris, France) 1:200 and the streptavidin-biotin-peroxydase complex.

For immunofluorescence staining, sections were incubated overnight with primary antibodies in concentrations of twice the dilutions used above and revealed with a secondary antibody conjugated to FITC (Eurobio, France). The second incubation was performed with either Cy3-conjugated anti-GFAP or anti-NG2, which was detected by a secondary antibody conjugated to Alexa Fluor 555 (Molecular Probes, Interchim, Montluçon, France). To label immature OLs (O4) and microglia on a section, anti-O4 was revealed in red (Alexa Fluor 555) and microglia in green (FITC-conjugated IB4). Immunofluorescence combined with TUNEL staining was done with fluorescein-TUNEL labeling first, followed by immunocytochemistry. Controls obtained by omitting the primary antibody were run for each immunocytochemical experiment. Tissues sections were photographed with a photometric device camera (Leica, DFC 300 FX) interfaced with IM50 imaging software.

Myelin Basic Protein (MBP) quantification. A computerized video-camera-based image-analysis system (with NIH image software) was used for densitometry measurements, as reported (Liu *et al.*, 2002). All available sections (at least four per bregma and per brain) were analyzed. Only sections with obvious technical artifacts related to the staining procedure were excluded. Unaltered TIFF images in the external capsule and cingulum were digitized, segmented (using a consistent arbitrary threshold −50%), and binarized (black versus white); then total black pixels per 2500 pixel-square were counted, and average values were calculated. Because there were interassay differences in the intensity of MBP immunostaining, no attempt was made to compare absolute OD values among experimental groups. OD were expressed as ratio of left-to-right measurements; for each brain sample, $(L:R)_{MBP}$ of pixels was calculated.

Data analysis. The data are expressed as the mean ± S.E.M. and values were compared using the Mann-Whitney nonparametric test. A value of $p<0.05$ was considered to be significant.

RESULTS

Histopathology

We have previously reported that the left MCA electrocoagulation associated to the left MCA occlusion leads to ipsilateral cerebral infarction (Renolleau *et al.*, 1998). Examination of cresyl violet-stained sections showed a well-delineated cortical infarct lesion with a mean infarct volume of $58 \pm 4$ mm$^3$ (mean $\pm$ S.E.M.) at 48 hours after reperfusion (Fig. 1 A and C). When brains were examined 2 weeks after the ischemic onset, the ipsilateral hemisphere exhibited a large cavity in the full-thickness of the frontoparietal cortex (Fig. 1 B and D).

Neonatal ischemia causes white matter injury

In the developing rat brain during the second and third postnatal weeks, there is a substantial increase in MBP immunostaining. Following ischemia in P7 rat pups, a decrease in ipsilateral MBP labeling was observed, compared to the contralateral side (Fig. 2). To evaluate overall trends in the recovery of MBP density, MBP immunostaining in each cerebral hemisphere was estimated by the computerized image-analysis-based semiquantitative method for measurement of MBP immunostaining recently described by Liu et al. (Liu *et al.*, 2002). Table 1 presents calculated $(L:R)_{MBP}$ values for each brain in the external capsule and cingulum (at bregma 0.2 and –2.5 mm, respectively). During the first 12 hours of recovery, a rapid decline in $(L:R)_{MBP}$ value was observed in the external capsule at bregma 0.2 mm. This decrease was less pronounced at bregma –2.5 mm ($p< 0.05$, versus bregma 0.2 mm), a level with a reduced infarct area as compared to the anterior level (Ducrocq *et al.*, 2000). Two days post-ischemia, ratio values were approximately half of age-matched sham (0.96 ± 0.06) or control (1.03 ± 0.14) values at both levels. No recovery in MBP immunostaining was detected at time periods up to two weeks after ischemia. In the cingulum (bregma 0.2 mm), no significant difference was detected in the $(L:R)_{MBP}$ values up to 15 days of recovery. In contrast, there was a substantial difference in $(L:R)_{MBP}$ values at 48 and 72 hours ($p<0.05$, Mann-Whitney test) at bregma –2.5 mm, indicating more MBP immunostaining in the ipsilateral than the contralateral hemisphere (see Fig. 2 D, E). Between 72 hours and 7 days post-ischemia, $(L:R)_{MBP}$ values declined to a mean of 1.16 which remained stable at 15 days after ischemia (Table 1).

Cell death in the white matter following neonatal ischemia

No TUNEL labeling was detected in the sham-operated or control rat pup brains at P7 and P8. In contrast, chromatin condensation and fragmentation was observed in the cortex and external capsule (Fig. 3 A). Quantification of TUNEL-positive nuclei in the external capsule demonstrated that most cell death occurred during the first 24 hours of reperfusion (Table 2). The number of TUNEL-positive nuclei decreased thereafter until 72 hours of recovery. Identification of these TUNEL-positive cells by double immunofluorescence indicated that 4.9 ± 0.4 cells/0.1 mm$^2$ (Table 2) were astrocytes (GFAP labeling, Fig. 3B) and 2.5 ± 0.2 cells/0.1 mm$^2$ were mature oligodendrocytes (MBP immunoreactivity, Fig. 3C). With increased reperfusion time, astrocytes exhibited short and thick processes (becoming reactive) and some of them had TUNEL-positive nuclei (Table 2). Within 12 hours after ischemic insult, numerous pyknotic O4-positive OLs exhibiting condensation of the cell body and fragmentation of the process arbor, and labeling of both plasma membrane and cytoplasm were detected in the external capsule (Fig. 3D) and were still seen at 24 and 48 hours post-ischemia (Table 2). Double TUNEL-O4 immunostaining (to detect dying immature OLs) was not performed because Triton-X100/acetic acid treatment used for the TUNEL assay induced the solubilization of the O4 antigen. In the cingulum, only a few TUNEL-positive nuclei were detected during the first 24 hours after reperfusion. A small number of pyknotic O4-positive OLs and GFAP-TUNEL astrocytes were observed (Table 3). No TUNEL-positive nuclei were seen thereafter.

Neonatal ischemia triggers transient cell proliferation

To investigate the increase in myelin seen at bregma −2.5 mm, we first evaluated cell proliferation by the presence of Ki-67 marker, a nuclear protein expressed in dividing cells for the entire duration of their mitotic process and expressed in all mammalian species (Scholzen & Gerdes, 2000; Kee *et al.*, 2002) in the cingulum above the dorsolateral border of the ipsilateral hippocampal SVZ. As shown in Figure 4, a significant increase in Ki-67 marker was observed between 24 ($91 \pm 8$ cells/0.1 mm$^2$, $p<0.05$) and 72 hours ($312 \pm 12$ cells/0.1 mm$^2$, $p<0.01$) post-ischemia (Fig. 4 E). This increase was followed by a decrease at 7 days post-ischemia. However, at this time point Ki-67 labeling was high at the border of the cavity close to the dorsolateral SVZ (Fig. 4 D). A similar increase in Ki-67 labeling was also observed in the contralateral cingulum, but the number of Ki-67 cells at 48 and 72 hours post-ischemia were only 50 % of those on the ipsilateral side. The first peak of Ki-67-positive cells was followed by a second one occurring between 7 and 21 days, with a peak of 148 ± 11 cells/0.1 mm$^2$ (p<0.05) at 15 days after ischemia (Fig. 4 E).

We next determined the nature of dividing cells in the ipsilateral hemisphere. By P7 most of the OL progenitors in the corpus callosum have matured to immature OLs and no NG2 (early OLs) labeling was observed. In contrast, a very significant increase of NG2-positive cells was observed between 48 and 72 (peak of 219.4 ± 19.2 cells/0.1 mm$^2$) hours after reperfusion (Fig. 5) followed by a decrease at 7 days in the ipsilateral cingulum. In the contralateral side, a slight increase of NG2-positive cells was also found and their number was not significantly different between ipsi- and contralateral sides at 24 and 48 hours (ratio of 1.2 and 1.1, respectively). This ratio increased up to 6.8 at 72 hours, then decreased to 1.8 at 7 days post-ischemia. No NG2 immunolabeling was detected at 10 and 15 days post-ischemia. At these ages, numerous early OLs have matured to immature O4-labeled OLs and a 4-to-5 fold increase in O4-positive cells was found in the ipsilateral compared to the contralateral cingulum (see Fig.7 A-B).

Using double immunofluorescence, we determined that around 60-70 and 35 % of the Ki-67-positive cells were astrocytes and early OLs, respectively at 48 hours (Fig. 6). At 72 hours of recovery most of double-labeled cells (around 85 %) were positive for NG2 (Fig. 6).

Microglia activation in the white matter

*Griffonea Simplicifolia* I isolectin B4 and OX-42 are markers of reactive microglia. These markers do not distinguish between blood-derived macrophages and resident/activated microglia, but both cells displayed a different morphology, with stout processes for ameboid microglia or a round shape and no processes for macrophages (Marty *et al.*, 1991; Benjelloun *et al.*, 1999). During the second week of life in control pups, resting microglia with ramified processes were detected throughout the corpus callosum. Following ischemia with reperfusion in P7 rat pups, microglial activation was observed in the cingulum. An increase greater than 2 to 3-fold was observed between 24 and 72 hours in activated microglia with an ameboid shape, as compared to age-matched control pups (62.5 ± 7.5 versus 23.3 ± 1.7 cells/ mm$^2$, p<0.001 at 48 hours), followed by a decrease. A similar increase was also observed contralaterally (not shown). In the cortical infarct and sub-cortical white matter a prominent increase in IB4-positive macrophages/microglia was observed between 72 hours and 2 weeks after reperfusion. These macrophages were present at the border of the cyst and were shown to engulf immature O4-positive OLs (Figure 7 C, D) leading to an ongoing death process, which contributes to the cyst growth.

DISCUSSION

The present findings demonstrate immature OL and astrocytic death and microglial activation in the white matter following transient unilateral focal ischemia with reperfusion in P7 rats. The developing brain also responds to ischemic injury by generating new precursors. In addition, the sustained presence of microglia-macrophages at the immediate border of the growing grey and white matter cavity may prevent OL maturation. These data suggest that inflammation via microglia may additionally contribute to immature OL injury in hypoxic-ischemic white matter injury.

In this neonatal stroke model, blood recanalization in the left common carotid artery was induced by the release of the clip. At this time, a reperfusion phase occurs via anastomoses between anterior, middle and posterior arteries in the MCA territory as demonstrated by black ink perfusion. This reperfusion has been detected before brain infarction was visible (pale zone). Indeed, first signs of cell death appear in a few scattered TUNEL-positive nuclei in the fronto-parietal cortex at 4-6 hours of recovery with a maximum of cell death occurring between 24 and 96 hours (Renolleau et al., 1998). A similar cerebral blood flow is restored to control values immediately upon return to normoxic conditions following hypoxia-ischemia as demonstrated by the indicator diffusion technique using iodo[$^{14}$C]-antipyrine (Mujsce et al., 1990).

As previously reported in the "Rice-Vannucci" model of HI (Liu et al., 2002), ischemia with reperfusion in P7 rat pups results in progressively widespread disruption of MBP immunostaining within 24 hours after reperfusion in the external capsule, which persists for two weeks. The initial loss of MBP immunostaining at 12 hours post-ischemia reflects oligodendroglial dysfunction, but only a few oligodendrocytes were detected with TUNEL-positive nuclei, suggesting that mature myelinated OL were normal. In contrast, numerous immature pyknotic OLs were observed, indicating their vulnerability to HI as previously reported in neonatal mice (Skoff *et al.*, 2001) and rats (Jelinski *et al.*, 1999, Back, 2001 #41, Ness, 2001 #31; Back *et al.*, 2001; Ness *et al.*, 2001). During OL lineage maturation *in vivo*, Back et al. (Back *et al.*, 2001) demonstrated that late OL progenitors (O4$^+$/O1$^-$) are selectively susceptible to HI compared to earlier (NG2$^+$, O4$^-$) or later (O4$^+$/O1$^+$) stages in P7 rats. However, NG2-positive pro-OLs have been reported to show DNA fragmentation and activated caspase-3 protein in the periventricular white matter after perinatal hypoxia-ischemia in P7 rats (Ness *et al.*, 2001), and such vulnerability may also account in our stroke model. Both deaths of early and immature OLs contribute to the initial loss of MBP immunostaining. No loss of MBP immunostaining was detected in the cingulum at bregma 0.2 mm but a significant trend in increase MBP labeling was observed at bregma -2.5 mm. A substantial restoration of MBP immunostaining has been reported in mildly lesioned P7 rats after 1.5 h hypoxia compared to a more severe HI insult (2.5 h hypoxia) (Liu *et al.*, 2002). These data and our findings suggest either regeneration of intact myelin sheaths by surviving mature OLs able to myelinate surviving axons in remote areas not directly affected by ischemic injury, or proliferation of new pro-OLs capable to mature into myelinating OLs. Taken together, our data suggest that regeneration may occur in area, which suffer mild ischemic insult (cingulum) as compared to more severe insult induced in the cortex and sub-cortical white matter.

Little is known about *in vivo* repair mechanisms following ischemia in the developing brain. Previous *in vitro* studies have observed that oligodendrocyte progenitors have the potential to differentiate into mature oligodendrocytes (Levine *et al.*, 2001), suggesting that the reaction of OL progenitors during the post-ischemic reperfusion period may represent tissue repair mechanisms such as the replenishment of OLs and the regeneration of myelin. Recent studies reported upregulation of OL progenitors associated with restoration of mature OLs and myelination in peri-infarct area in the adult (Tanaka *et al.*, 2003) and immature (Zaidi *et al.*, 2004) brain. Most of the studies used the cell proliferation-specific marker 5'-bromodeoxyuridine (BrdU) to demonstrate increased neurogenesis after cerebral ischemia in the adult rat and mouse hippocampus (see (Kokaia & Lindvall, 2003) for review). Using the same paradigm we recently reported an increased BrdU-positive cell number in the border (penumbra) of the cortical infarct 7 days following ischemia in P7 rats (Benjelloun *et al.*, 2003b). Using antibodies against Ki-67, a reliable marker of mitosis, we found a significant increase of Ki-67-positive nuclei between 48 and 72 hours in the ipsilateral cingulum above the dorsolateral hippocampal SVZ compared to control and to 7 days after ischemia. At this time point of recovery, Ki-67 labeling was seen in the deeper cortical layer surrounding the infarct and in white matter tracts, suggesting migration of newly generated cells from dorsolateral SVZ toward lesioned cortex and white matter. Most of the Ki-67-positive nuclei represent proliferative NG2-positive pro-OLs and astrocytes during the first week of recovery in response neonatal stroke, as previously reported following HI (Levison *et al.*, 2003). Upregulation of NG2 expression has also been observed following ischemia-reperfusion in adult rat (Tanaka *et al.*, 2003). Recently, BrdU+/CA-II+ (carbonic-anhydrase II, labeling cell bodies of OLs) following HI in P7 rat brain were detected 12-14 and 21-22 days after HI injury (Zaidi *et al.*, 2004). It is difficult to compare this study and ours because of the differences between the 2 models used and the timing of the study. Interestingly, the developing brain is able to respond to ischemic insult by upregulating NG2 expression, which may be an adaptative mechanism attempting to remyelinate brain tissue. In contrast, HI depleted neural stem cells and OLs progenitors from the SVZ within hours of the insult, and the SVZ remained poorly repopulated at 21 days of recovery (Levison *et al.*, 2001). The discrepancy between these results may be related to the severity of HI injury, and taken together our data suggested that ischemia-reperfusion in P7 rats is able to induce repair processes. In addition, we still observed Ki-67-positive cells at 15 days post-ischemia. Among these cells a few were double stained by BrdU and glutamate decarboxylase ($GAD_{67}$) and those labeled by nestin displayed neuronal or glial morphology (Benjelloun *et al.*, 2003b).

During development, microglial cells have an important role, in eliminating neurons and glial cells that are destined to die (Milligan *et al.*, 1991). In the control rat, a variation in the number of resident microglia was observed between P7 and P9 (not shown), a time window for which programmed cell death is nearly finished. In our model, a significant increase of ameboid microglia occurred between 24 and 72 hours post-ischemia, during the window of vulnerability to selective HI white matter injury, as compared to controls. In the P7 rat, microglia have been implicated in the pathophysiology of neonatal injuries with a transient increase in microglial antigens between 24 hours and 3-4 days post HI in P7 rats (McRae *et al.*, 1995), and after excitotoxic lesions of the developing murine periventricular white matter (Tahraoui *et al.*, 2001). Here, we demonstrated that i) early NG2-positive OLs matured into O4-positive OLs in the cingulum and were also present at the vicinity of the cyst, and ii) increased macrophage/microglia phagocytosed O4-positive OLs, suggesting that the presence of activated microglia interferes with the maturation of immature OLs leading to impaired remyelination.

In conclusion, our data demonstrate that the reperfusion phase following ischemia in the developing rat brain leads to a sustained inflammatory response, which prevents repair processes. Understanding such events in acute, subacute, and chronic white matter lesions in human hypoxic-ischemic encephalopathy will be essential for choosing effective therapeutic targets.

Acknowledgments : We are grateful to Professor F. Gold (Hopital Trousseau) for support, and to P. Bouquet (UMR-CNRS 7102) and V. Cochois (Université R. Descartes) for technical assistance, and to Ann Lohof (UMR-CNRS 7102) for editorial assistance.

FIGURE LEGENDS

Figure 1 : Neonatal ischemia histopathological changes of the left cerebral hemisphere. A – D : Representative Cresyl violet-stained coronal sections from animals killed at 48 hours (A, C) and 15-17 days (B, D). Note the large ill-defined pale area at 48 hours corresponding to the large cavity in the full-thickness of the frontoparietal cortex 2 weeks post-ischemia.

Figure 2 : Neonatal ischemia induced changes in MBP immunostaining. A: P8 control rat at the level of the anterior commissure (bregma 0.2 mm). MBP immunostaining is barely discernible in the external capsule and corpus callosum. B–C: MBP immunostaining is reduced in the ipsilateral external capsule (B) compared to the contralateral side (C), 24 hours post-ischemia. Note the marked loss of MBP-immunostained axonal processes. D–E: MBP immunostaining was not significantly different in the cingulum between ipsi- and contralatral hemisphere at 24 h. The two boxes in A represent areas in which MBP immunostaining was quantified.  Bar represents 1 mm (A) and 25 $\mu$m (B-E).

Figure 3 : Demonstration of cell death in the ipsilateral cortex and white matter at 24 hours of reperfusion. A: Dying cells in the cortical infarct and external capsule (ec) demonstrated by DNA fragmentation and apoptotic bodies (arrows) using the TUNEL assay. V, VI = layer V and layer VI of the cortex; ec = external capsule. B and C: Double-stained sections GFAP-TUNEL (B) and MBP-TUNEL (C) at 24 hours after reperfusion. B: An astrocyte (red, white arrow) with a TUNEL-positive nucleus (green, white arrow). C: Desorganized myelin fibers (red) and an oligodendrocyte (myelin

labeled in red) with a TUNEL-positive nucleus (white arrow) D: Example of a pyknotic immature oligodendrocyte with no discernible processes labeled with O4 antigen. Scale bar represents 50 $\mu$m (A) and 25 $\mu$m (double immunostaining).

Figure 4 : Neonatal ischemia triggers cell proliferation as demonstrated by Ki-67 immunoreactivity. Ki-67-positive nuclei were detected in the cingulum and dorsolateral border of the SVZ at 24 (A) and 72 (B) hours and 7 (C) days (n = 3 or 4 animals) post-ischemia. At 7 days of recovery, Ki-67-positive nuclei were mainly detected in the subcortical layer of the cortical cystic infarct (D, star) but less in the cingulum (C). Bar represents 100 $\mu$m (A-D). E: Quantification of Ki-67 nuclei (box outlined in A). By 48 and 72 hours after reperfusion, a significant increase in the density of Ki-67-positive nuclei occurred in the ipsilateral (**$p<0.02$) and contralateral (*$p<0.05$) cingulum and dorsolateral border to the hippocampal SVZ.

Figure 5 : Neonatal ischemia triggers activation of early OL progenitors as demonstrated by NG2 immunostaining. A–B: Demonstration of NG2 upregulation in the ipsilateral (A) as compared to the contralateral (B) corpus callosum, at 72 hours of recovery. The star indicates the cortical lesion and arrows the enlarged view at the level of the ipsilateral (C) and contralateral (D) cingulum. E: Quantification of NG2-positive cells (n= 3 or 4 animals per time point) in the cingulum. By 72 hours of reperfusion, a significant increase in the density of NG2-positive nuclei (**$p<0.02$) occurred in the ipsilateral cingulum at the hippocampal level (Bregma –2.5 mm),

compared to the contralateral side. Enlarged panel represents 2 NG2-positive OLs. Bar represents 500 (A-B), 100 (C-D) and 25 (enlarged panel) $\mu$m.

Figure 6 : Identification of proliferating cells at 48 and 72 hours of recovery. Double immunofluorescence staining to detect new astrocytes (labeled by Ki-67 and GFAP) at 48 hours and newly formed oligodendrocytes (OLs, labeled by Ki-67 and NG2) at 72 hours at the level of the cingulum. Bar represents 25 $\mu$m.

Figure 7 : Immunostaining against O4 antigen 15 days following ischemia and reperfusion. A–B: O4-positive cells in the contralateral (A) and ipsilateral (B) SVZ, illustrating an increased number of immature OLs in the latter. C–D: Double stained sections with anti-O4 (red) and isolectin B4 (green). Note that several microglia/macrophage engulfed O4-positive pyknotic immature OLs at the border of the cavity (star). Bar represents 75 (A, B), 150 (C, D) and 100 $\mu$m (enlarged panel).

Table 1 : Densitometric analysis of MBP immunostaining after ischemia and reperfusion in P7 rat pups. MBP immunostaining in each cerebral hemisphere was estimated by computerized densitometry (see Materials and methods); for each sample, $(L:R)_{MBP}$ of pixels was calculated at the level of the external capsule and cingulum and at two levels (bregma 0.2 and –2.5 mm) in a minimum of 3-4 animals at each time point of recovery.

| (L:R)MBP | 12 h | 24 h | 48 h | 72 h | 7 days |
|---|---|---|---|---|---|
| | Bregma 0.2 mm | | | | |
| External capsule | $0.64 \pm 0.1^{\#, \S}$ | $0.67 \pm 0.08$ | $0.59 \pm 0.06$ | $0.61 \pm 0.12$ | $0.48 \pm 0.08$ |
| Cingulum | $1.05 \pm 0.19$ | $1.01 \pm 0.25$ | $0.97 \pm 0.11$ | $1.10 \pm 0.12$ | $0.95 \pm 0.08$ |
| | Bregma – 2.5 mm | | | | |
| External capsule | $0.81 \pm 0.07$ | $0.60 \pm 0.17$ | $0.61 \pm 0.12$ | $0.52 \pm 0.02$ | $0.57 \pm 0.08$ |
| Cingulum | $1.01 \pm 0.16$ | $1.19 \pm 0.12$ | $1.27 \pm 0.12^*$ | $1.30 \pm 0.18^*$ | $1.16 \pm 0.14$ |

- * $p < 0.05$, Mann-Whitney versus 12 h time point

- [#, §] p<0.05, Mann-Whitney versus sham (0.96 ± 0.06) and 12 h-value (bregma –2.5 mm), respectively.

<u>Table 2</u> : Quantification of cell demise in the white matter (external capsule and cingulum, at bregma –2.5 mm) after reperfusion after ischemia in P7 rat pups. Cell death was demonstrated by the TUNEL assay and dying astrocytes or mature oligodendrocytes were detected by the double staining (TUNEL and GFAP or MBP immunostaining, respectively). Pyknotic immature OLs were quantified on their morphology (condensation of the cell body and fragmentation of the process arbor) and labeling of both plasma membrane and cytoplasm with the O4 antibody. Data are expressed in mean ± SEM (cells per 0.1 mm$^2$, n= 3 or 4 at each time point of recovery).

|  | External capsule | | | Cingulum |
|---|---|---|---|---|
|  | 24 h | 48 h | 72 h | 24 h |
| TUNEL-positive nuclei | 10.5 ± 1.2 | 6.8 ± 1.5 | 1.6 ± 0.3 | 3.3 ± 0.8 |
| GFAP$^+$/TUNEL$^+$ Cells | 4.9 ± 0.4 | 2.6 ± 0.5 | 1.3 ± 0.5 | 1.8 ± 0.7 |
| MBP$^+$/TUNEL$^+$ Cells | 2.5 ± 0.2 | 1.2 ± 0.4 | 0 | 0 |
| Pyknotic O4$^+$ Cells | 9.8 ± 3.7 | 5.3 ± 0.9 | 0 | 1.7 ± 0.5 |

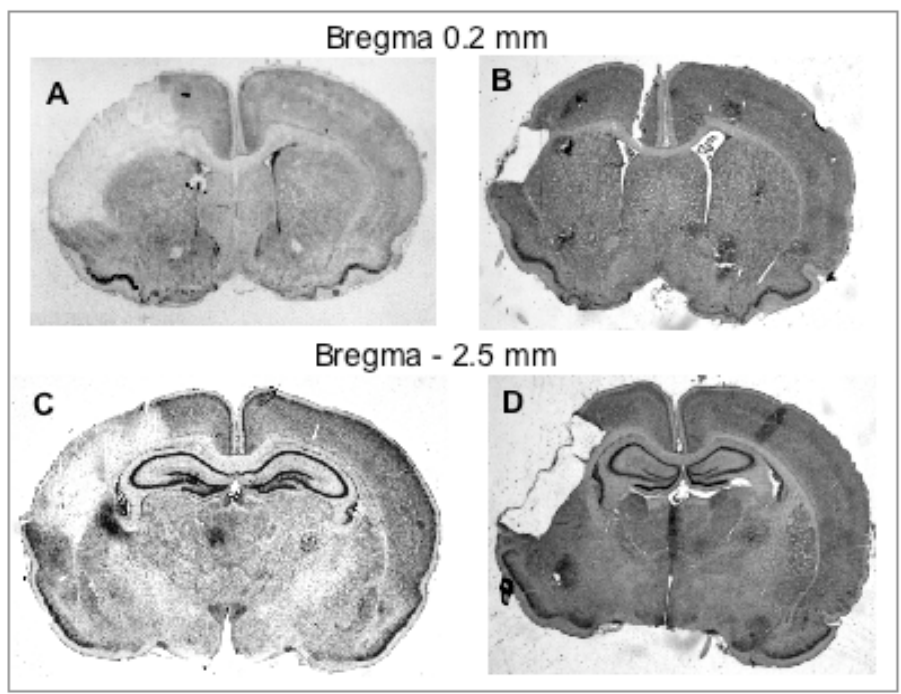

Figure 1

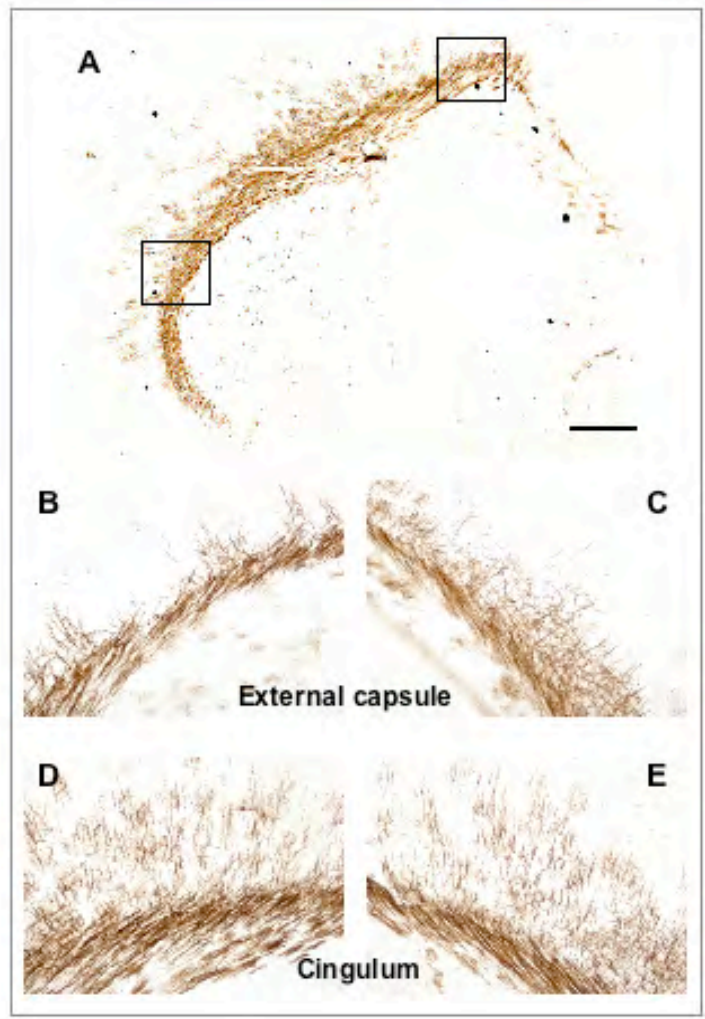

Figure 2

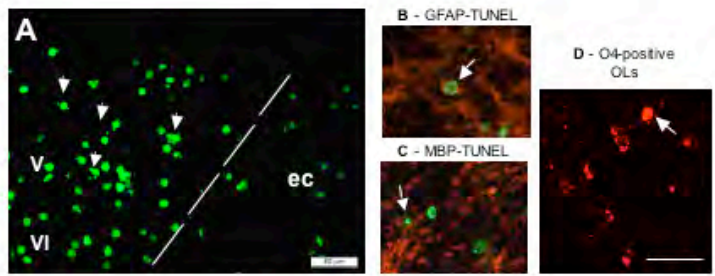

Figure 3

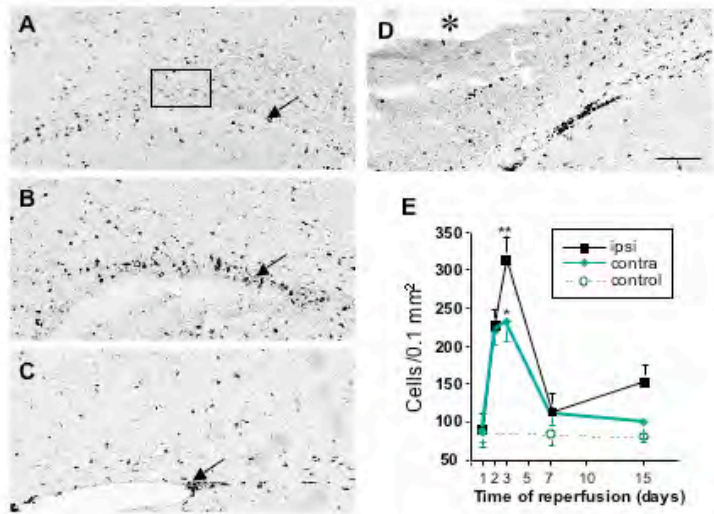

Figure 4

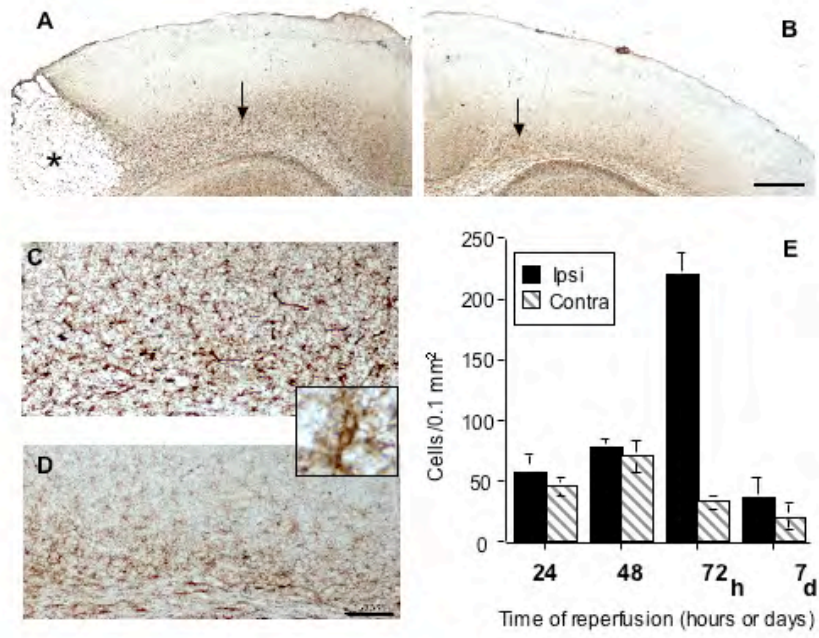

Figure 5

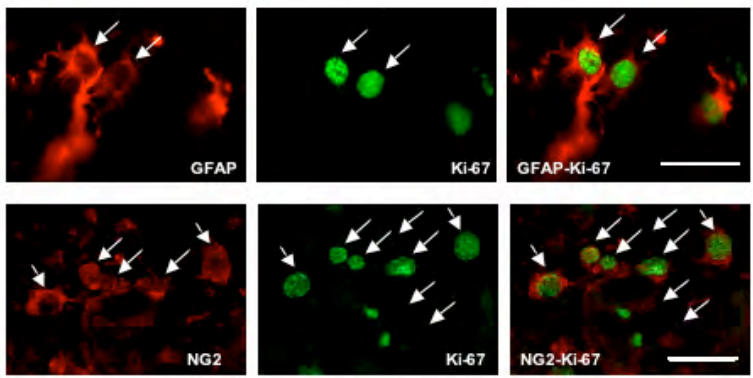

Figure 6

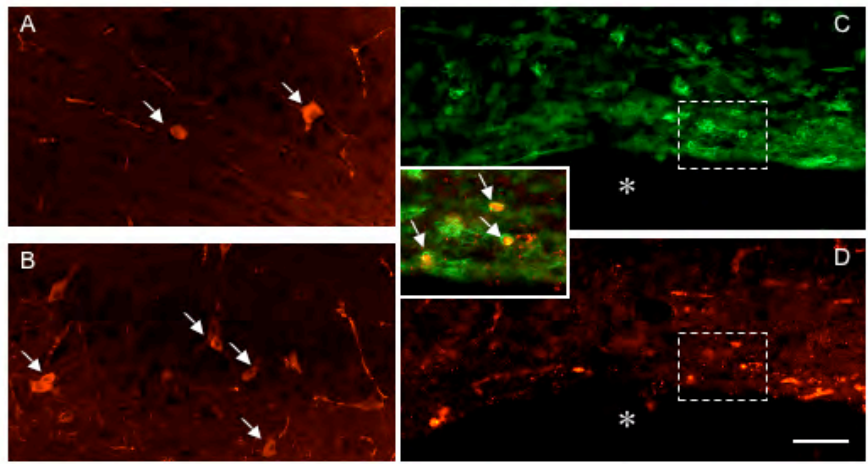

Figure 7